# Contact Lens with Moiré patterns for High-Precision Eye Tracking

I.M. Fradkin[1], R.V. Kirtaev[1], M.S. Mironov[1], D.V. Grudinin[1], A.A. Marchenko[1], M.M. Chugunova[1], V.R. Solovei[1], A.V. Syuy[1], A.A. Vyshnevyy[1], I.P. Radko[1], A.V. Arsenin[1], V.S. Volkov[1,*]

[1]Emerging Technologies Research Center, XPANCEO, Dubai Investment Park First, Dubai, United Arab Emirates

*Correspondence should be addressed to e-mail: vsv@xpanceo.com

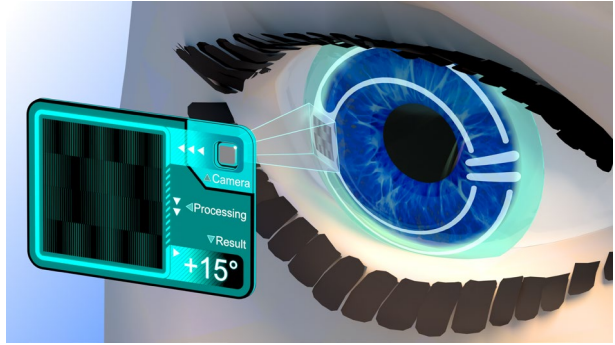

## Abstract

Eye tracking is a key technology for human–computer interaction, particularly crucial in augmented reality (AR) and virtual reality (VR) systems. We propose a novel eye-tracking approach based on incorporating passive eye-tracking modules into contact lenses. These modules comprise two superimposed gratings separated by a narrow gap. The overlapped gratings produce moiré pattern, while the spatial separation between them results in parallax effect, namely, pattern transformation upon variations in viewing angle, which enables accurate angular measurements. This method is insensitive to ambient lighting conditions and requires neither scale and color bars nor perspective corrections. Using this approach, we have experimentally measured lens orientation with angular resolution exceeding 0.3°, which is satisfactory for gaze detection in most AR/VR applications. Furthermore, the proposed technological platform holds a potential for many-fold enhancement in measurement precision.

## Introduction

Humans acquire most environmental information through visual perception. In this context, eye tracking is a crucial technology for assessing human attention, concentration, and human–machine interaction in general. Notably, basic eye-tracking techniques emerged as early as the 19th century, revealing fundamental aspects of vision such as saccades and fixation points[1], which are now considered foundational knowledge. Currently, eye-tracking technologies serve diverse purposes, from scientific research[2] and medical applications[3–6] to commercial uses in marketing[7,8], safety monitoring[9], and assistive technologies[10,11]. Eye tracking is already used in VR/AR applications[12] and will likely play an even more crucial role in the future generation of devices.

While multiple eye-tracking techniques exist and are being developed[13–21], most current applications utilize computer-vision-based approaches[22–28]. These solutions leverage recent advances in camera module miniaturization, powerful computational capabilities, and novel analytical tools like neural networks, enabling precise and fast measurements. However, several significant challenges still remain, which leaves room for further development. First, an infrared illumination is required to improve optical contrast of images. Yet even in this case, environmental light sources produce a noticeable background and complicate the measurements. Second, continuous monitoring of eye motion is a very energy-consuming process, limiting most portable devices to several hours of autonomous operation.

Contact lenses represent a promising platform for vision-related devices. Recent developments include integrated intraocular-pressure sensors[29–41] and tear-based biosensors[42–44]. Beyond medical applications, contact lenses are increasingly viewed as a potential platform for future AR/VR projection systems[45,46] and other applications[47–49]. Accurate image projection requires continuous eye-position monitoring, particularly when a contact lens (with an embedded projection system) exhibits rotational or translational (sliding along the eyeball) freedom relative to the eye. Current lens-integrated tracking solutions include embedded coils[50–53], photodetectors[54,55], and even microscopic accelerometers and magnetometers[56,57]. While these approaches can exceed the precision of standard computer vision methods, they often present limitations such as requiring eye anesthesia, bulky design, or complex fabrication processes. This also makes the problem of lens-assisted eye tracking relevant.

This work presents a simple, cost-efficient label embedded into a contact lens for high-precision eye tracking that requires a non-specialized external camera module[58]. The tracking label is a passive optical element comprising two overlapped gratings that exhibit a moiré pattern highly sensitive to the viewing angle due to the parallax. Relative shift of moiré patterns enables precise tracking of lens rotation, achieving angular resolution of approximately 0.3 degrees with a potential for further improvement.

# Results

## Label operating principle

In this study, we propose a label embedded in a contact lens (see Fig. 1a) that streamlines eye-tracking procedures while enhancing precision and reducing computational demands for image analysis. The label is a passive optical element, whose appearance, when observed by an external camera module, varies substantially with a change of the viewing angle. This enables precise evaluation of the angle between the camera module and the tracking label that is fixed within a contact lens. While wearing contact lenses for eye tracking might initially seem cumbersome or unnecessary, it is worth noting that more than 140 million people[59] already use contact lenses. Furthermore, high-precision eye tracking is primarily intended for lens-integrated AR systems, making the integration of an auxiliary tracking module a natural extension.

Computer vision researchers frequently encounter challenges related to perspective correction in image analysis. One of the ways to address this issue is to employ reference markers with specific geometric patterns. A priori knowledge of their geometry allows determining the orientation of these markers relative to the observation camera. However, eye-tracking applications impose unique constraints: tracking markers must be extremely small while maintaining exceptional precision. Therefore, this study focuses on enhancing measurement precision by developing tracking markers whose appearance is extremely sensitive to variations of the observation angle.

One of the effective methods for increasing viewing angle sensitivity is through parallax. However, the limited thickness of a contact lens (typically several hundred micrometers) restricts potential parallax displacement. Even a tilt of several degrees produces a parallax shift of only a few micrometers. To detect such minute displacements, the reference elements on the tracking marker must be similarly microscopic. Since these elements are too small for direct observation without microscopy, we utilize the moiré effect to achieve observable large-scale variations. As illustrated in Fig. 1c, when two identical gratings of a small period $p$ unresolvable by an imaging system are perfectly overlapped without relative shift or rotation, they transmit light homogeneously without forming a specific pattern. However, light transmission depends on the observation angle. Thus, when observed normally to the surface [Fig. 1c(i)], the transmission is maximum. When the observation angle $\Theta = \arctan\frac{p}{2H} \approx \frac{p}{2H}$, where $H$ is the gap between the gratings, the transmission is minimum, because the slits of one grating are overlapped by the lines of the other grating [Fig. 1c(ii)]. At approximately double this angle, $\Theta = \arctan\frac{p}{H} \approx \frac{p}{H}$, the transmission is maximum again [Fig. 1c(iii)]. This can be illustrated by a transmission diagram [Fig. 1c(iv)] that is constant for varying grating coordinate $x$, but changes periodically with the observation angle $\Theta$. These intensity variations enable us to determine the viewing angle within the angular period $a^\Theta = \frac{p}{H}$ (in the paraxial approximation). Although the brightness of such a structure is determined solely by the observation angle, different points within the structure may be observed at slightly different angles, which can introduce transmission gradients. This implies that, strictly speaking, the measured intensity should be associated with the observation angle of the corresponding points, rather than with that of the entire eye-tracking module.

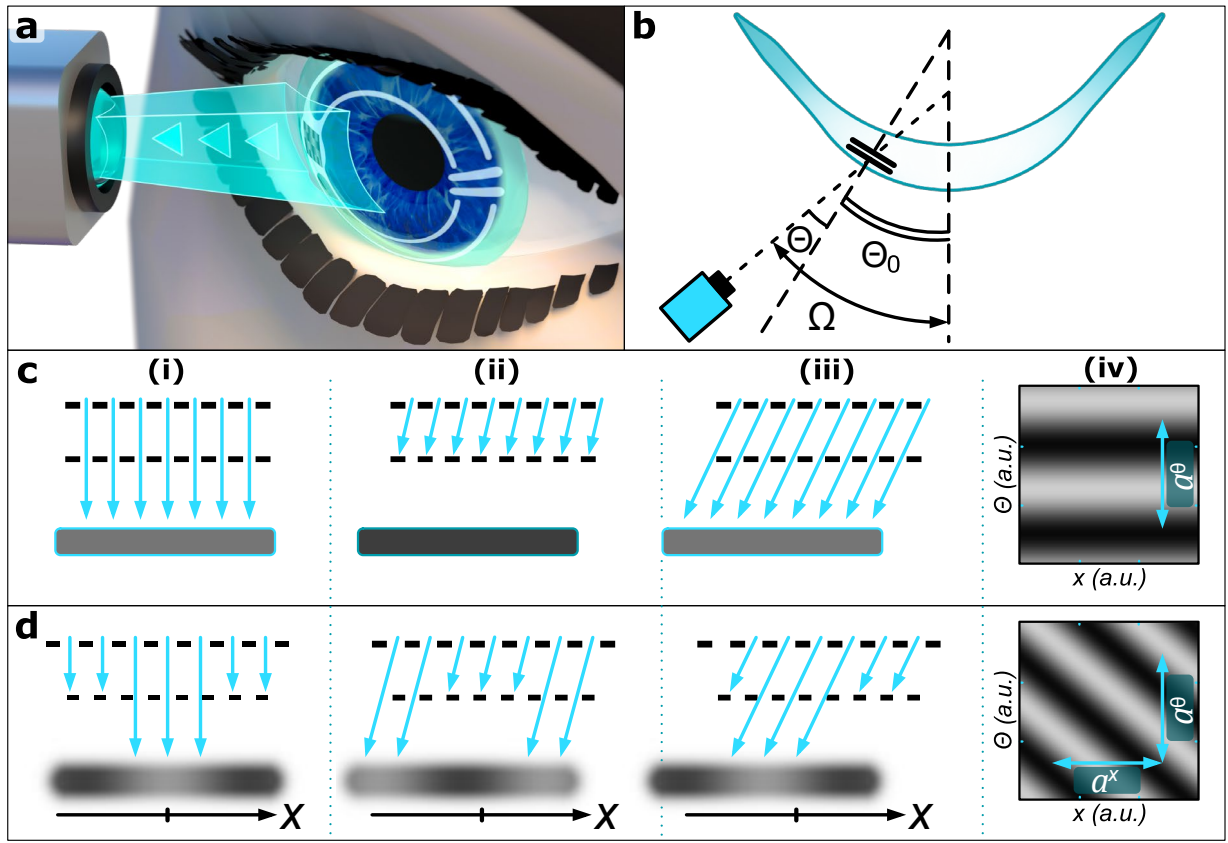

**Figure 1. a** Contact lens with an integrated eye-tracking label, observed using an external camera module to measure lens orientation. **b** Contact lens with the eye-tracking label cross-section illustrating the camera observation angle relative to the label normal ($\Theta$) and relative to the lens axis ($\Omega$). **c** (i-iii) Schematics of the parallax effect in a stack of two identical gratings.

(iv) Stack transmission diagram (in zero-angular-size approximation) that is independent of the spatial coordinate *x*, but varies periodically with the observation angle Θ. **d** (i-iii) Schematics and (iv) transmission diagram demonstrating that a slight mismatch in grating periods introduces moiré fringes with a finite spatial period while maintaining the original angular periodicity.

Bilayer grating structure provides angular sensitivity observable at macroscopic scale, though several challenges remain for its practical implementation. First, the structure brightness used to determine the observation angle depends significantly on ambient lighting conditions, camera parameters, and other external factors. Second, after a single angle measurement using a specific point of the label, identifying the same point in subsequent measurements can be problematic due to potential confusion with neighboring points, which inherently limits measurement accuracy. Furthermore, the same label may appear differently when observed from various angles and distances, which complicates the point-matching process. Third, the angular dependence of the label brightness is periodic, potentially restricting the range of unambiguous measurements. Below we discuss our technological solutions to address these challenges.

To help matching the label intensity with the observation angle, we can incorporate a reference pattern, akin to a color bar in color maps, that experiences identical lighting conditions. Notably, the bilayer label itself can serve as its own reference scale if it contains at least one complete intensity period. This arrangement allows simultaneous observation of maximum and minimum intensities that can be used to calibrate label brightness.

When designing such a label using two identical gratings, the minimum required label size can be estimated as $da^\Theta \approx d\frac{p}{H}$, where $d$ is the camera-to-lens distance. This estimation assumes that the approximation of small-angular-size label does not hold and the label appears to host the full angular period $a^\Theta$ of intensity because its opposite sides are observed at different angles. Conservative estimations for typical dimensions ($H \sim 200$ μm, $p \sim 30$ μm, $d \sim 10$ cm) indicate that the label size should become comparable with the size of a contact lens (15 mm) making it impractical for integration.

However, we can significantly reduce the label size by introducing a slight mismatch between the periods of the two stacked gratings. As illustrated in Fig. 1d, such a structure produces periodic fringes called moiré pattern even in the small-angular-size approximation when the opposite sides of the label are observed at approximately the same angle. Introduced modification results in a finite spatial period of moiré fringes $a^x = p\frac{p}{\Delta p}$ (with $\Delta p$ being the difference in periods of the gratings), while the angular period $a^\Theta$ remains the same (see Fig. 1d(iv)) since it depends solely on the ratio of the (slightly modified) grating period to the gap between gratings. For a sufficiently small label, we can assume uniform observation angle across its area (zero-angular-size approximation), so the fringes period is fully determined by the grating period mismatch, whereas their phase corresponds to the parallax-induced shift, i.e., the phase depends on the observation angle.

This approach, initially developed to obtain a reference intensity map, enables measurement of fringes phase and period rather than intensity at specific points. This method does not only ensure independence of measurements on lighting conditions, but also collects information from multiple points, substantially improving the signal-to-noise ratio and reducing measurement errors.

Even after successfully determining the fringes phase and deriving from it the observation angle, it is crucial to keep in mind that the phase should be determined relative to some origin that is the same across all measurements. While edges, corners, or the label frame can serve as references, relying on just a few of them for phase evaluation ultimately limits the overall precision of measurements by the accuracy of determining the location of these points. Since locating an individual point is highly susceptible to noise and limited by the observation-camera pixel size, it is better to measure the fringes phase relative to a complex object composed of a very large number of points. Another periodic structure can be a good choice for this purpose. Thus, we propose integrating at least two elements that exhibit different moiré patterns on a single eye-tracking label. If the moiré patterns shift differently relative to the label (or, equivalently, shift relative to each other) with the change of the observation angle, it is sufficient to measure their relative phase instead of the absolute phase of a single moiré pattern. We realize multiple different moiré patterns by adjusting the period mismatch between upper and lower gratings, keeping their average period almost unaffected.

To extend the angular range of eye tracking beyond the moiré pattern angular period $a^\Theta$, one can implement continuous eye tracking. Since the angle variations are continuous, the moiré fringes phase can be determined unambiguously. Also, if moiré structures have different angular periods, their least common period might be large enough to cover the whole range of interest.

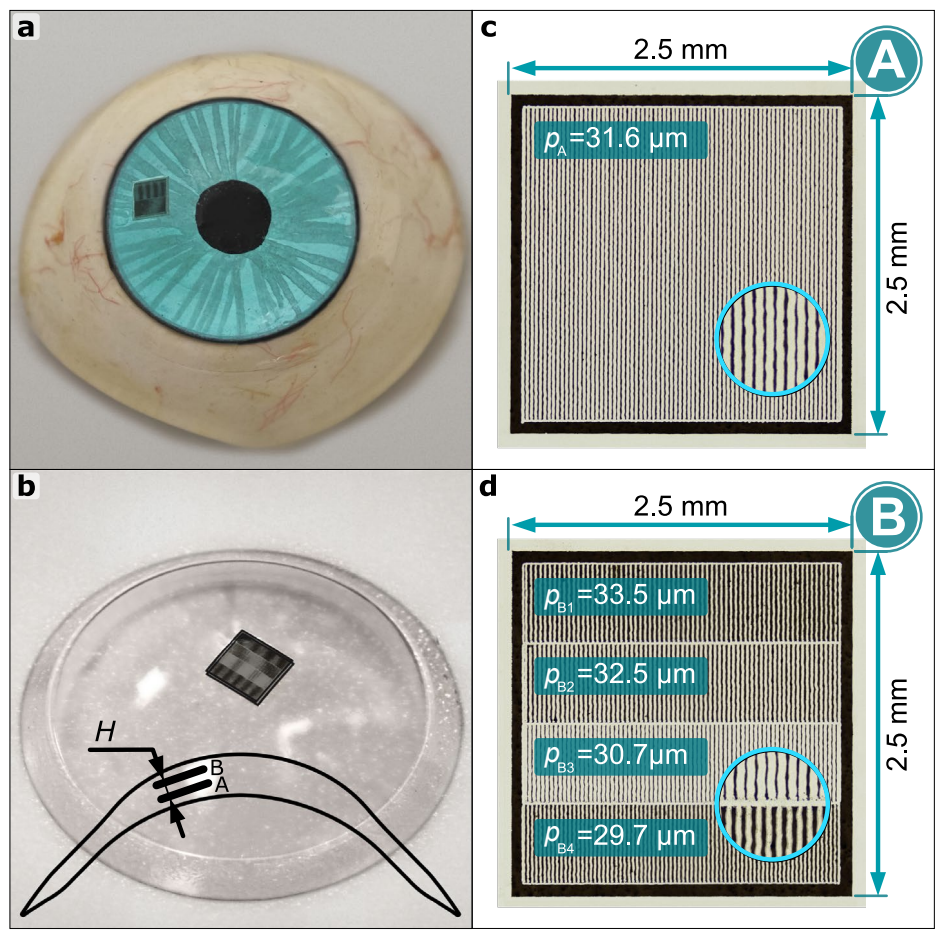

**Figure 2. a** Photograph of a contact lens with an integrated eye-tracking label placed atop of an eye model. **b** Contact lens with a label comprising four elements exhibiting distinct moiré patterns explored in the article. Inset: Schematics of the contact lens cross-section showing the eye-tracking label, where top grating B is separated from bottom grating A by a distance $H \approx$

250μm . **c,d** Microphotographs of gratings A and B, respectively. Grating B consists of four subgratings with different periods $p_B$ denoted on the figure, while grating A (reference) has a single period $p_A$. Circular insets demonstrate the enlarged areas of the gratings.

## Label design and characterization

For practical implementation, we developed a contact lens with an integrated bilayer grating, as shown in Fig. 2a-2b. In this work, we analyze the sample shown in Fig. 2(b): grating A (bottom) with the designed period $p_A = 31.6$ μm serves as a reference (see Fig. 2c), while grating B (top) consists of four subgratings with the designed periods $p_B = 29.7, 30.7, 32.5,$ and $33.5$ μm, corresponding to mismatches of ±2.9% and ±6.0% relative to the reference grating A (see Fig. 2d). These dimensions result in moiré fringes periods of approximately 31.6 μm/0.029 ≈ 1.1 mm for the two central areas of the label and 31.6 μm/0.06 ≈ 0.5 mm for the other two external areas, realizing hereby four areas with distinct moiré patterns. The difference in fringe periods can be seen in Fig. 2b. The actual dimensions of the gratings and moiré periods deviate slightly from the designed values due to deformation, thermal shrinkage, and other manufacturing-related effects that occur during and after the lens fabrication. The gap between gratings, which is difficult to control precisely, can be reliably determined from the calibration of the labels and was $H \approx 250$ μm in our experiment, as demonstrated below. Notably, although the spatial periods $a^x$ of the structures differ, they all share nearly identical angular periods $a^\Theta \approx 10°$, indicating similar sensitivity to changes of the observation angle.

Due to the small grating periods, its internal structure is not resolvable and only moiré fringes are visible on photographs (see Figs. 2a,b). The intensity of a moiré pattern can be described by a simple harmonic function:

$$I(x) \propto \cos\left(\tilde{k}(x - \tilde{x})\right), \qquad (1)$$

where $\tilde{k} = 2\pi/a$ represents the grating vector of the corresponding moiré pattern and $\tilde{x}$ denotes its shift from a reference position. In Supplementary Information (Note 1), we demonstrate that the grating vector $\tilde{k} = |k_A - k_B|$ is almost independent of $\Theta$, while the shift $\tilde{x}$ is proportional to $\Theta$ for small observation angles (when $\tan\Theta \approx \Theta$).

Thus, it is possible to extract parameters of each moiré pattern from a photograph and determine the observation angle $\Theta$ from the shift value $\tilde{x}$. However, as mentioned above, to obtain reliable results, the label on each snapshot must be analyzed in the same coordinate system tied to this label and scaled together with the photograph, which is challenging and may limit the overall accuracy of this approach.

Consequently, it is more practical to express the observation angle $\Theta$ through a relative shift, i.e., the shift of one moiré pattern relative to another. Following this approach, below we consider the 1st and 4th moiré patterns on the label. First, the observation angle $\Theta$ in air and that in the material of the contact lens (in our experiment, polydimethylsiloxane, PDMS) are related by the Snell law $\sin\Theta = n^{lens}\sin\Theta^{lens}$. Second, simple derivations (see Supplementary information, Note 1) provide a relation between $\Theta^{lens}$ angle and parameters of the observed moiré patterns:

$$\tan\Theta^{lens} = \frac{p_B}{H}\frac{\tilde{x}_4 - \tilde{x}_1}{a_1 + a_4} + C, \qquad (2)$$

where $C$ is a constant determined by the microscopic parameters of the gratings A and B producing the moiré patterns. Crucially, the observation angle $\Theta$ is determined by the dimensionless ratio of the patterns relative shift to the sum of their periods, which is independent of location and size of the label on the snapshot. Furthermore, even if the image of the label is distorted due to large viewing angles or some optical aberrations these distortions to find the normalized relative shift between gratings. This eliminates the need for scale bars, further simplifying the fabrication process.

While the observation angle $\Theta$ is defined with respect to the label normal, at this stage we focus only on angle changes rather than absolute angles. Therefore, we measure only the angle change relative to a reference orientation, $\Omega = \Theta + \Theta_0$ (see Fig. 1b). For simplicity, the camera reference orientation is aligned with the contact lens axis. In this configuration, $\Theta_0 \approx 5°$ represents the relatively small angle between the label normal and the contact lens axis (see Fig. 1b). Given the relatively narrow range of observation angle variation $\Omega \in [-15°, 15°]$, we can apply the paraxial approximation, $\tan \Theta^{\text{lens}} \approx \Theta/n^{\text{lens}}$, in Eq. (2) and consider a linear angular dependence of relative moiré shift.

Experimental measurements were performed using the setup schematically depicted in Fig. 3a. A contact lens equipped with an eye-tracking label was mounted on a rotating stage and imaged from a distance of approximately 40 cm. Figure 3b(i) shows photographs of the eye-tracking label captured at four observation angles: -10°, -9°, -8°, and -7°. With the continuous change of the angle, the two upper moiré patterns shift in the direction opposite to that of the two bottom moiré patterns. This difference in directions is due to the opposite sign of the period mismatch for the corresponding grating B relative to the reference grating A (Fig. 2c,d).

To enhance visual clarity, shifts measured by our algorithm (Supplementary Information, Note 2) are marked on the photographs as circles and accompanied by horizontal bars scaled proportionally to their magnitudes, displayed alongside for easier comparison. The proportionality between the shift value and the moiré period confirms that their ratio is a critical parameter. A separate column (ii) displays computer renders of the designed structure, which reproduce the observed effects and corroborate our interpretation. In this column, circles and bars correspond to theoretically predicted shifts. As shown, the general behavior of the experimental sample aligns with the model. However, in experiment, the two bottom moiré patterns (3rd and 4th) exhibit larger shifts than the two upper ones (1st and 2nd), despite theoretical predictions of almost equal displacements. This discrepancy stems likely from deformations of the gratings during encapsulation into the contact lens.

In total, 31 images of the eye-tracking label were captured at a 1° rotation increment for analysis. Relative shifts (Fig. 3c) and periods (Fig. 3d) of the moiré patterns were extracted by analyzing local maxima in the Fourier space (Supplementary Information Note 2). Importantly, all values were measured in pixels of the corresponding acquired images since maintaining constant magnification across different snapshots is unnecessary. The only requirement for image analysis is a common coordinate system for all moiré patterns within the same image, which is inherently satisfied. Consequently, fluctuations in shifts and periods (Fig. 3c,3d) should not be directly interpreted as errors but rather as deviations in the choice of coordinate systems for processing of different images. The measurement error arises from the first term in Eq. (2) (moiré patterns relative shift normalized by the sum of their periods) and would be considered below.

As it is evident from Fig. 3c, the pattern shifts exhibit a near-linear dependence on observation angle. Measurement of the third moiré pattern shift shows the highest error due to its period being comparable to the label's total size. Meanwhile, moiré periods (Fig. 3d) remain nearly constant across the measurement range, indirectly confirming the same scaling across all snapshots.

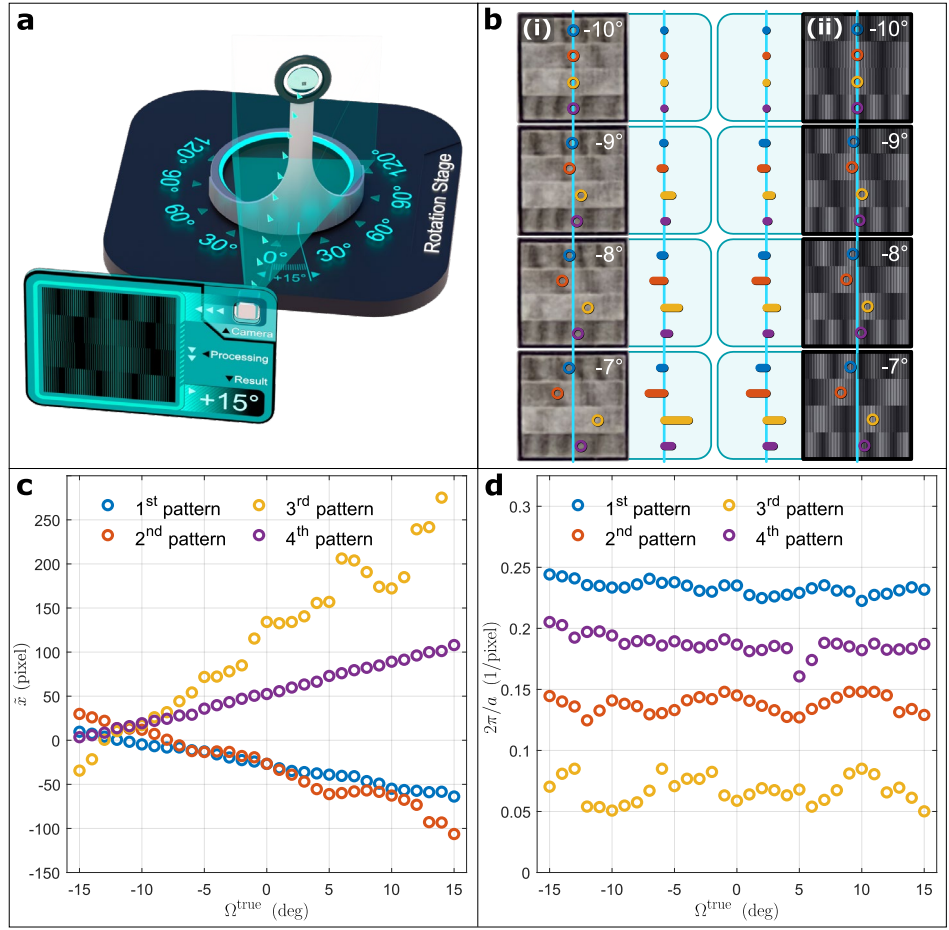

**Figure 3. a** Schematics of the experimental setup. A contact lens with an embedded eye-tracking label is mounted on a rotation stage for precise angle control and imaged by an external digital camera. **b** (i) Photographs of the label at four different observation angles. (ii) Computer renders of the designed structure at the same observation angles as in the experiment. Circles and horizontal bars indicate shifts of the moiré patterns. **c** Shifts of four moiré patterns (in pixels of the imaging camera) versus observation angle. The first two patterns (1st and 2nd) shift in the opposite direction compared to the last two (3rd and 4th). **d** Grating vector (in inverse pixels) of the four moiré patterns as a function of observation angle. Comparisons are valid only for grating vectors and shifts evaluated from the same photograph.

### Angle measurements and precision estimation

Using images of the moiré patterns, we analyzed the relationship between observation angle and normalized relative pattern shift according to Eq. (2). Figure 4a demonstrates this relationship derived from comparing two outermost moiré patterns (1st and 4th), which shift in opposite directions, have the smallest periods, and provide stable results. The observed linear dependence confirms suitability of the label for angle measurement. By fitting experimental data to a linear function, we can determine both the angular period $a^\Theta = \frac{p_A}{H} n^{\text{lens}} \approx 10.4°$ from the slope of the line fit and structure thickness $H = \frac{p_A}{a^\Theta} n^{\text{lens}} \approx \frac{31.6\,\mu m}{10.4°\frac{\pi}{180°}} 1.43 \approx 250\,\mu m$. We evaluate the measurement error, $\sigma_{41}^\Theta = 0.41°$, as the root mean square deviation of the true viewing angle $\Omega^{\text{true}}$ from the values predicted by linear fit $\Omega_{41}^{\text{est}}$. The typical error is approximately 1/25 of the label's angular period, demonstrating that the approach is suitable for detecting subtle moiré pattern shifts.

Current design shows potential for increasing precision of measurements through reducing the grating periods. Indeed, the smaller the grating period, the smaller the angular period of the moiré patterns, $a_\Theta$, which implies that the same relative shifts will be observed when rotating at a smaller angle. However, we can improve precision even with the existing label design. Using the four available moiré patterns, in total six pairs can be formed, four of which group those patterns that shift in opposite directions, which is optimal for signal-to-noise ratio. Fig. 4b compares estimations from these four pairs based on their measurement-data linear fits (detailed in Supplementary Information, Note 3). The measured angle deviations from true values often show different signs and uncorrelated behavior. Averaging data from the four moiré-pattern pairs yields error of $\sigma^\Theta = 0.28°$ (Fig. 4c), which is better than any pair considered separately. Fig. 4d illustrates deviations of the estimated angles from the actual with larger errors occurring at wider angles, suggesting enhanced precision of approximately 0.2° or better for a narrow angle range typical in eye-tracking applications.

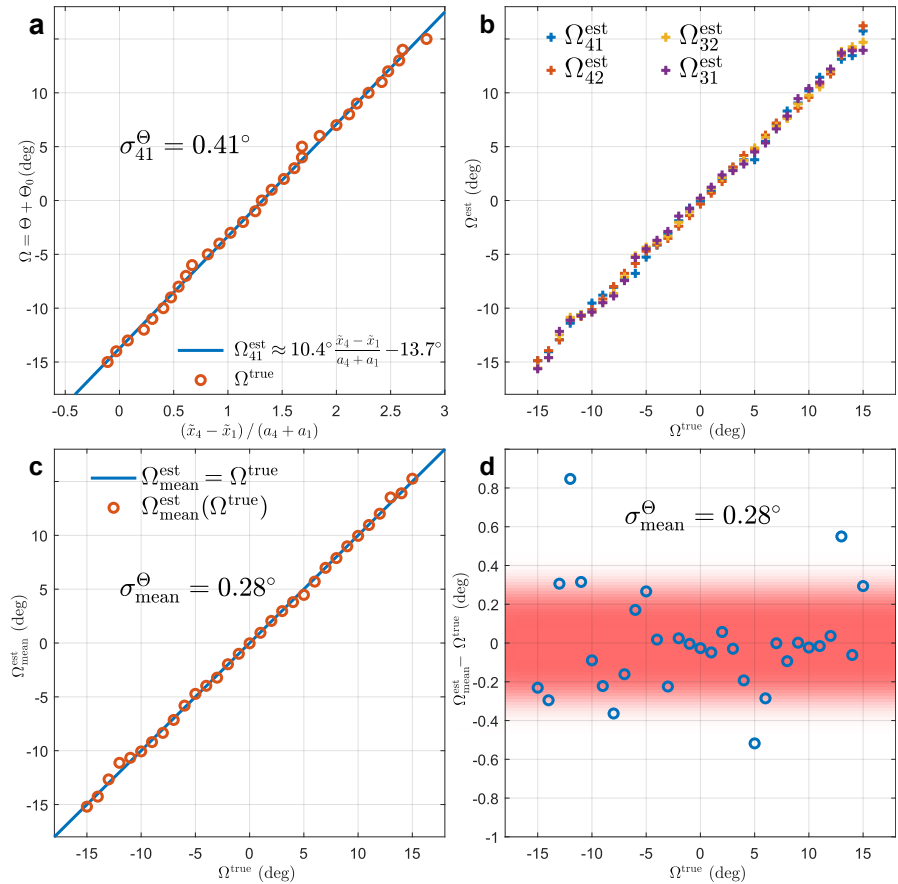

**Figure 4. a** Measurement data showing linear dependence of the normalized moiré pattern shifts (first and fourth patterns) on the change of the observation angle. **b** Angle estimates from four selected pattern pairs compared to the true values, revealing uncorrelated errors. **c** Angle estimation averaged across all four combinations, achieving 0.28° accuracy. **d** Error distribution, showing enhanced deviations at large angles. This suggests improved precision within a narrow angle range, important for AR applications.

While various improvements in data processing are possible – including weighted averaging of results from different structures, non-paraxial analysis, accounting for moiré pattern gradients, and enhanced noise suppression – the primary avenue for improvement lies in reducing grating periods to increase sensitivity. This approach could potentially improve precision by more than an order of magnitude, making the proposed method promising for eye and lens tracking in future augmented and virtual reality systems.

## Discussion

We have demonstrated a contact lens with an integrated passive optical parallax-based eye-tracking module that requires an external general-purpose camera observing the module. The system is highly sensitive to the observation angle variations, which facilitates simplified image processing. We utilized the moiré effect to form periodic patterns on the module. The patterns are characterized by their periods and relative shifts. These parameters, being illumination-independent and derived from a large number of data points, enhance signal-to-noise ratio and minimize relative errors.

Our innovative approach measures relative shifts between patterns rather than the absolute pattern displacement, eliminating the need for reference points and perspective distortion corrections. This methodology does not only streamline numerical analysis, but also significantly improves measurement precision. The implementation of multiple moiré patterns enables result averaging, further reducing uncorrelated random error contributions. All together, these measures enabled angular precision better than 0.3° – approximately 1/37 of the angular period of the parallax-sensitive moiré structure – validating the effectiveness of our methodology. Notably, enhanced precision (better than 0.2°) was achieved within a ±10° around the center of the angle range, meeting typical AR/VR application requirements.

Beyond high precision, our solution offers practical advantages for large-scale implementation. The eye-tracking module features a straightforward, relatively easy to manufacture design that maintains high precision even with rather large (tens of micrometers) structural elements. The system requires a standard camera module – already present in most current solutions – and does not need auxiliary illumination, allowing seamless integration with the existing AR/VR devices without hardware modifications. Our system uses image processing via computationally efficient algorithms, which enables continuous eye tracking without excessive power consumption. Furthermore, there is a potential for precision enhancement, primarily through reduction of grating periods, which can increase measurement accuracy at least several-fold.

## Materials and Methods

### Fabrication

The gratings composing the eye-tracking label were made of a 200-μm-thick layer of polydimethylsiloxane elastomer (PDMS, Sylgard 184) with opaque lines formed by a thin layer of black ink (black tattoo ink from Dynamic Color Co, USA; a carbon black solution in isopropanol with water). First, Sylgard 184 was prepared by thoroughly mixing of the pre-polymer base with crosslinking curing agent as per datasheet. Then, the mixture was degassed in a centrifuge and poured into a syringe. The syringe was installed into the pump of a slot-die coater (Slot-Die Coater from Ossila Ltd, UK) for subsequent coating onto standard microscope slides (76×26×1 mm); the thickness of the coating layer was controlled by built-in micrometres. After deposition, PDMS films were cured in an oven (OVEN 125 basic drying oven from IKA-Werke GmbH, Germany) with a schedule according to Sylgard datasheet. In order to make the PDMS surface hydrophilic for coating with the ink, the substrates were treated in air plasma (ATTO plasma system from Diener electronic GmbH, Germany). Afterwards, a thin layer of the black tattoo ink was spin-casted onto the PDMS surface (Advanced Spin Coater from Ossila Ltd, UK) and dried on a hotplate. Gratings of the required design (distributed on the substrate with sufficient separation for subsequent handling) were directly written on the ink layer using the laser engraving machine (Muse UV Galvo Laser engraver with 355 nm DPSS source from Full Spectrum Laser LLC, USA). The gratings were cut out from the PDMS layer with a sharp scalpel, leaving the outer frame of blank PDMS, and transferred to blank carrier glass slides. The reference grating A was transferred with the ink side facing up, while the complementary grating B was placed with ink facing the glass slide. Further assembly employed PDMS-to-PDMS bonding and was performed on the HB16 wire-bonding machine with H80 pick&place option (all from TPT Wire Bonder GmbH, Germany) that had a mouse-type manipulator for in-plane XY direction, manual rotating stage, and a motorized Z axis equipped with a vacuum pick-up tool. The slide with grating B was placed and clamped on the stage, while the substrate carrying grating A was aligned with grating B and brought in contact with a help of the built-in microscope, XYZ controls, and a vacuum tool. This resulted in a sandwiched structure with the following order of layers (from bottom to top): glass slide, ink layer of grating B, 200-µm PDMS film, ink layer of grating A, 200-µm PDMS film, glass slide. Then the carrier glass was detached leaving a 400-µm-thick eye-tracking module. At the last step, the module was encapsulated into a mold-cast contact lens also made of Sylgard 184 PDMS material. For this, the module was dipped into PDMS and placed in the female mold at the desired location, after which the polymer was partially cured in the oven. Then, the mold was fully filled with PDMS, covered with the male mold, and placed into the oven for complete curing.

## Measurements

A contact lens with an integrated eye-tracking label was mounted on a motorized rotation stage approximately 400 mm away from a digital camera. Initial placement was arranged such that both the camera's optical axis and the lens's symmetry axis were collinear, with the sample image forming near the centre of the camera's digital sensor. The contact lens was then rotated in a ±15° range with a 1° step around an axis that would correspond to the eye rotation axis. At each lens position, the eye-tracking label was photographed by the stationary camera using consistent settings. Regular room lighting conditions were sufficient to acquire high-quality digital images.

The rear camera of a Samsung SM-S928B smartphone was used for this purpose with the following settings: 18.6 mm focal length, f/3.4 relative aperture, 1/50 s exposure time, and ISO sensitivity 200–300. Each photograph was taken with resolution 3060 x 480 pixels, and at 24-bit colour depth. To ensure precise and smooth rotation, we used Standa 8MR151 motorized rotating stage.

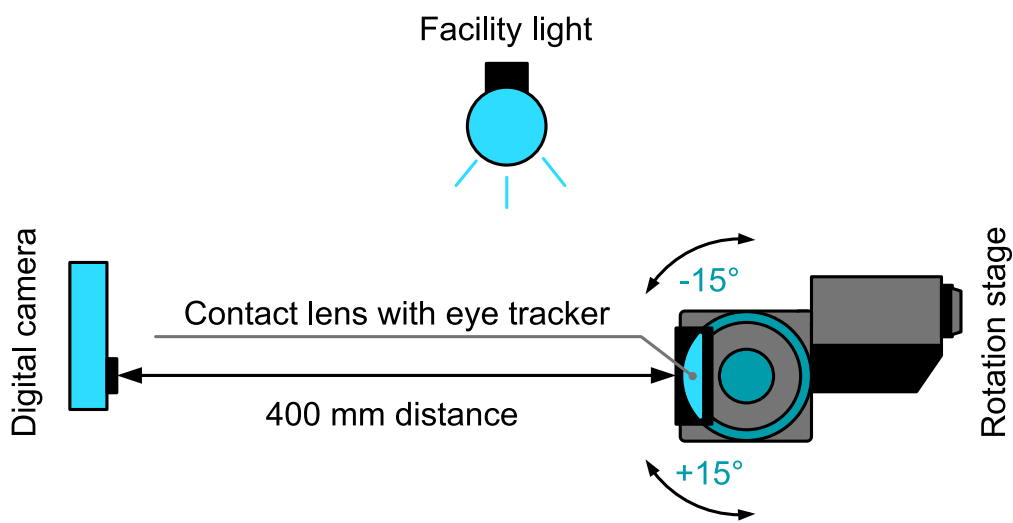

**Figure 5.** Schematics of the experimental setup. The contact lens is mounted on a rotation stage (±15° range) and imaged by a digital camera at ~40 cm distance under external illumination.

## Author Contributions

R.V.K. and M.S.M fabricated samples; A.A.M. and R.V.K. provided the measurements; I.M.F and D.V.G. performed data processing; I.M.F, I.P.R and A.A.V. prepared draft of the manuscript; M.M.C. and A.V.S. provided patent/literature search contributed in direction of the study; A.V.A. and V.S.V supervised the study. All authors reviewed and edited the paper. All authors contributed to the discussions and commented on the paper.

## Competing Interests

The authors declare no competing interests except for the patent WO2025018917A1 associated with this work and held by XPANCEO RESEARCH ON NATURAL SCIENCE L.L.C.

## Data Availability

The datasets generated and analyzed during the current study are available from the corresponding author upon reasonable request.

# Supplementary Information for

# Contact Lens with Moiré patterns for High-Precision Eye Tracking

I.M. Fradkin[1], R.V. Kirtaev[1], M.S. Mironov[1], D.V. Grudinin[1], A.A. Marchenko[1], M.M. Chugunova[1], V.R. Solovei[1], A.V. Syuy[1], A.A. Vyshnevyy[1], I.P. Radko[1], A.V. Arsenin[1], V.S. Volkov[1,*]

[1]*Emerging Technologies Research Center, XPANCEO, Dubai Investment Park First, Dubai, United Arab Emirates*

*\*Correspondence should be addressed to e-mail: vsv@xpanceo.com*

## Note 1: Theoretical background

In this section, we analyze the mechanism of moiré fringes formation in detail. Understanding the dependence of moiré pattern parameters on the viewing angle is critical for reconstructing this angle from observed changes in the pattern. As previously noted in the article, the PDMS/air interface introduces refraction governed by Snell's law. For simplicity, we therefore relate the moiré pattern parameters to the angle of light propagation within the polymer, denoted as $\Theta^{\text{lens}}$, rather than in air. Additionally, we assume the camera or observer is positioned at a sufficient distance from the module to ensure all its parts are viewed at the same angle.

The grayscale intensity of visible fringes can be approximated by multiplying the transparencies of the upper and lower gratings, which are shifted relative to each other by $H\tan\Theta^{\text{lens}}$. To simplify calculations, we model each grating's transparency as harmonic rather than the true piecewise-constant periodic function. This approximation does not affect the periods or phases of the structures, which are our primary focus. The intensity is then expressed as:

$$I^{\text{microscopic}}(x) \propto \sin\left(k_A(x + H\tan\Theta^{\text{lens}}) + \varphi_A\right) \cdot \sin(k_B x + \varphi_B) =$$
$$\tfrac{1}{2}\cos\left((k_A - k_B)x + k_A H\tan\Theta^{\text{lens}} + \varphi_A - \varphi_B\right) - \tfrac{1}{2}\cos\left((k_A + k_B)x + k_A H\tan\Theta^{\text{lens}} + \varphi_A + \varphi_B\right).$$
(S1)

For a distant camera, only the first, low-frequency term corresponding to the moiré fringes is observable. As shown, the fringe period remains constant and depends solely on the mismatch between the upper and lower grating wave numbers ($k_A$ and $k_B$), while the phase varies with the angle. Introducing simplified notation:

$$I(x) \propto \tfrac{1}{2}\cos\left(\tilde{k}(x - \tilde{x})\right), \qquad (S2)$$

where $\tilde{k} = |k_A - k_B|$, $\tilde{x} = \dfrac{k_A H \tan\Theta^{\text{lens}}}{k_B - k_A} - \dfrac{\varphi_B - \varphi_A}{k_B - k_A}$. The direction of the fringe shift is determined by the relative grating periods. For Structures 1 and 2, where the lower grating has a smaller period than the upper one ($a_A < a_{B1/B2}$, hence $k_A > k_{B1/B2}$), an increase in $\Theta$ results in a negative shift of $\tilde{x}$. Conversely, the opposite behavior occurs for structures 3 and 4.

While absolute shifts could theoretically determine the angle, practical limitations in measuring absolute displacements necessitate analyzing relative shifts. Here, for an example, we consider the relative shift between the first and fourth gratings:

$$\tilde{x}_4 - \tilde{x}_1 = \left(\frac{1}{k_{B4}-k_A} - \frac{1}{k_{B1}-k_A}\right) k_A H \tan \Theta^{lens} - \frac{\varphi_{B4}-\varphi_A}{k_{B4}-k_A} + \frac{\varphi_{B1}-\varphi_A}{k_{B1}-k_A}. \quad (S3)$$

Given $k_{B4} > k_A > k_{B1}$ we define $k_{B1} - k_A = -\frac{2\pi}{a_1}$ and $k_{B4} - k_A = \frac{2\pi}{a_4}$. Substituting these into Eq. (S3) yields the angle expression:

$$\tan \Theta^{lens} = \frac{p_A}{H} \frac{\tilde{x}_4 - \tilde{x}_1}{a_1 + a_4} + \frac{p_A}{H} \frac{a_4(\varphi_{B4}-\varphi_A) + a_1(\varphi_{B1}-\varphi_A)}{a_1 + a_4} \frac{1}{2\pi}. \quad (S4)$$

The constant term can also be rewritten as:

$$\tan \Theta^{lens} = \frac{p_A}{H} \frac{\tilde{x}_4 - \tilde{x}_1}{a_1 + a_4} - \frac{p_A}{H} \frac{\tilde{x}_{0;4} - \tilde{x}_{0;1}}{a_1 + a_4} = \frac{p_A}{H} \frac{\Delta \tilde{x}_4 - \Delta \tilde{x}_1}{a_1 + a_2}, \quad (S5)$$

where $\tilde{x}_{0;4}$ and $\tilde{x}_{0;1}$ denote shifts of the patterns observed at normal viewing angle. The viewing angle $\Theta^{lens}$ thus depends on a dimensionless ratio of the shift to the sum of moiré periods, ensuring robustness against image scaling or perspective variations - a key advantage for practical applications.

## Note 2: Data processing algorithm

In the theoretical background section, we established a method to calculate the viewing angle using the periods and relative shifts of moiré patterns. However, extracting these parameters from photographs requires solving technical challenges. The workflow begins with preprocessing steps illustrated in Fig. S1. First, photographs are converted to grayscale, and the eye module is detected (Fig. S1a). Next, the module is cropped laterally to remove peripheral frames, and each moiré pattern is averaged across the strips (Fig. S1b) to enhance the signal-to-noise ratio compared to single-line-section intensity measurements. The resulting profiles of the moiré patterns are shown in Fig. S1c. Finally, the mean background is subtracted from each profile to get intensity functions $I(x)$ convenient for further analysis (Fig. S1d).

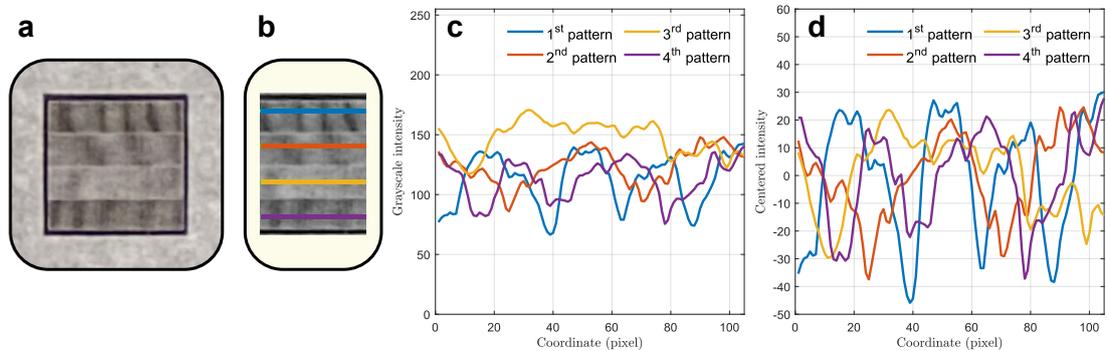

**Figure S1. a** Eye-tracking module in a photograph. **b** Cropped image with color strips marking grayscale intensity analysis regions. **c** Extracted grayscale moiré intensity profiles. **d** Background-subtracted moiré profiles used for data processing.

Further processing of $I(x)$ profiles employs Fourier analysis. The moiré wavenumber $\tilde{k}$ corresponds to the peak in the Fourier spectrum:

$$\tilde{k} = \underset{k}{\mathrm{argmax}} |F[I(x)]|(k),$$

while the relative shift is derived from the phase at this peak:

$$\tilde{x} = -\arg\left(F[I(x)](\tilde{k})\right)/\tilde{k}.$$

Two key limitations arise in practical implementation. First, discrete Fast Fourier Transform (FFT) restricts spectral resolution to $\delta\tilde{k} = \frac{2\pi}{D}$, where $D$ is the length of the sampled profile. This resolution limit, rooted in the uncertainty principle, theoretically bounds the precision of $\tilde{k}$ measurements. Second, finite sampling introduces discretization artifacts. However, the quasi-harmonic nature of moiré patterns enables precision beyond these constraints. A noise-free harmonic signal windowed by a rectangular function stably exhibits Fourier peaks closer to the true spatial frequency than predicted by the uncertainty principle.

To exploit this, we calculate true, continuous Fourier transform, which might be interpreted as a sort of clever interpolation of FFT spectrum between discrete grid points. Two equivalent strategies might be easily implemented.

1. Zero-padding: Extending the signal with null margins to artificially refine the FFT grid. It is highly efficient approach due to $O(N \log N)$ complexity of Fourier transform. Also it provides us the whole spectrum at once.
2. Convolution: Convolving the discrete FFT spectrum with the continuous Fourier transform of a rectangular window $\left(\propto \frac{\sin kD}{kD}\right)$ allows us to obtain the same result as the first approach. Though computationally intensive $(O(N^2))$ for the whole spectrum computation, this method allows to compute only limited number of points required efficiently localize the peak, which makes it fast as well. In practice we apply this approach.

Figure S2 compares discrete and continuous Fourier spectra for four moiré patterns. All spectra are zero-valued at $k = 0$, achieved by background subtraction. Continuous spectra (solid lines) pass strictly through FFT nodes (circles) but reveal peaks between grid points (see insets in Fig. S2 for enlarged images of peaks). This is especially important for large-period patterns (e.g., Pattern 3), where sub-grid shifts significantly affect accuracy. Multiple side peaks mainly arise from edge artifacts introduced by the rectangular cropping window. Continuous spectra enable consistent peak selection, avoiding ambiguities in discrete FFT analysis such as competing peaks of comparable amplitudes.

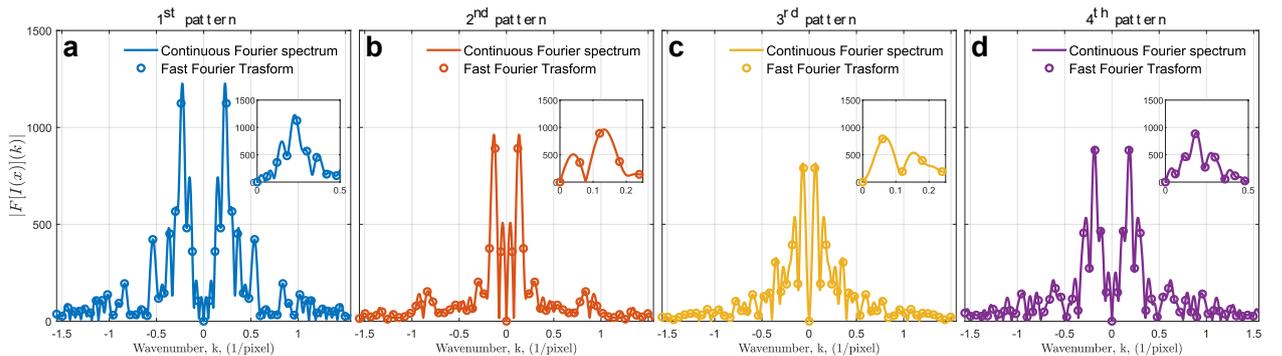

**Figure S2. a–d** Fourier spectra of four moiré patterns. Circles denote FFT data; solid lines represent continuous Fourier transforms. Insets: Enlarged peaks showing wavenumbers typically located between FFT grid nodes.

This approach enhances both period and shift determination, directly improving viewing angle calculations. While effective, further refinements using advanced signal and image processing techniques (e.g., optimized window functions or machine learning) could yield additional gains.

## Note 3: Angle measurement via different moiré patterns

As previously discussed, a single moiré pattern can be used to measure the viewing angle. However, using two patterns improves precision by enabling measurement of the relative shift rather than the absolute shift. With multiple distinct moiré patterns, the rotation angle can be determined from the relative shifts between any pair. The highest signal-to-noise ratio is expected for pairs of patterns shifting in opposite directions. Our design includes two left-shifting and two right-shifting patterns, yielding four combinations in total.

Figure S3 shows the normalized relative shift (divided by the moiré periods) as a function of rotation angles for all four combinations. All curves exhibit a linear relationship consistent with Eq. S4 under the paraxial approximation, confirming the structure's suitability for angle measurement. We quantify precision by calculating the root-mean-square deviation (RMSD) of true angles from linear-fit predictions. The minimal error, $\sigma^\Theta = 0.39°$, occurs for the pair combining the 4th and 2nd patterns, marking it as optimal. Other pairs achieve comparable precision, while combinations involving the noisier 3rd pattern perform worst.

Importantly, the slopes of all curves are almost the same, as predicted by Eq. S4. Minor deviations likely arise from the non-negligible mismatch between the periods of gratings A and B, which violates the assumption that their difference is small compared to individual periods. This mismatch also increases the least common period of the composite structure, potentially extending the unambiguous angular measurement range.

A critical limitation stems from the distinct constant terms (offsets) observed across pairs. These offsets reflect the random phases $\varphi_A$ and $\varphi_B$, which preclude advanced simultaneous fitting of all four patterns and necessitate pairwise analysis.

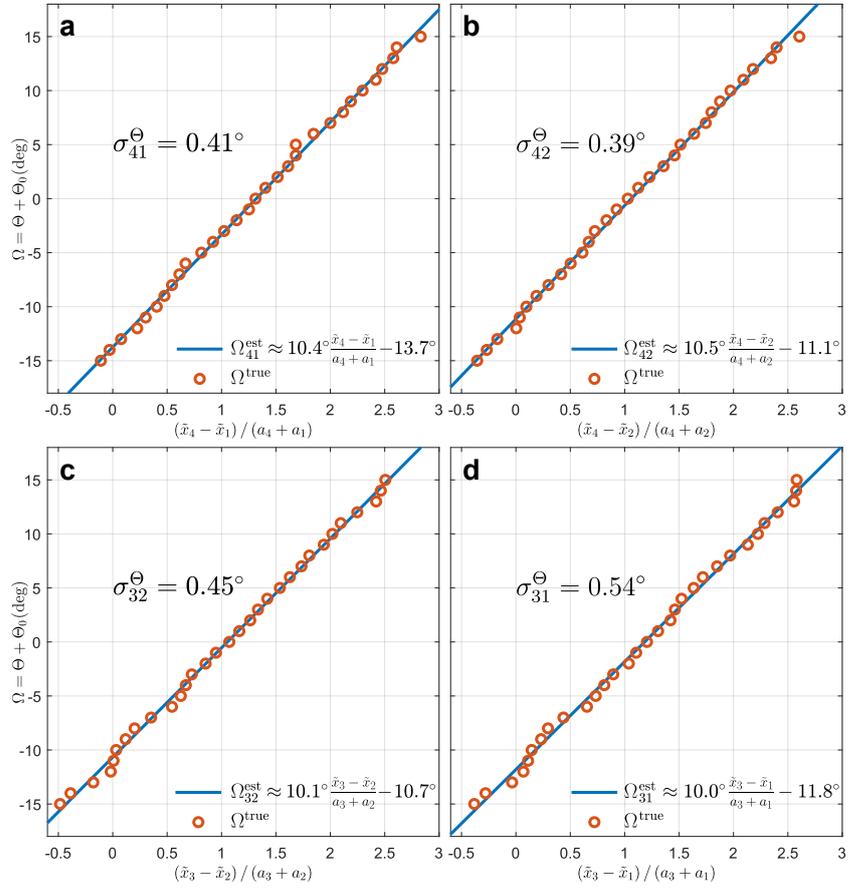

**Figure S3. a-d** Calibration graphs for relative viewing angle versus normalized moiré shifts across four pattern pairs. All pairs show comparable precision $\approx 0.4° - 0.5°$.